\documentclass[aps,pre,preprint,groupedaddress,showpacs]{revtex4-1}
\usepackage{wrapfig}
\usepackage{float}
\usepackage[dvips]{graphicx}
\usepackage{subfigure}
\usepackage{amsmath}
\begin{document}

% Use the \preprint command to place your local institutional report
% number in the upper righthand corner of the title page in preprint mode.
% Multiple \preprint commands are allowed.
% Use the 'preprintnumbers' class option to override journal defaults
% to display numbers if necessary
%\preprint{}

%Title of paper
\title{Growth states of catalytic reaction networks exhibiting energy metabolism}

% repeat the \author .. \affiliation  etc. as needed
% \email, \thanks, \homepage, \altaffiliation all apply to the current
% author. Explanatory text should go in the []'s, actual e-mail
% address or url should go in the {}'s for \email and \homepage.
% Please use the appropriate macro foreach each type of information

% \affiliation command applies to all authors since the last
% \affiliation command. The \affiliation command should follow the
% other information
% \affiliation can be followed by \email, \homepage, \thanks as well.
\author{Yohei Kondo}
\author{Kunihiko Kaneko}
%\email[]{Your e-mail address}
%\homepage[]{Your web page}
%\thanks{}
%\altaffiliation{}
\affiliation{Department of Basic Science, University of Tokyo, Komaba, Meguro-ku,
Tokyo 153-8902, Japan}

%Collaboration name if desired (requires use of superscriptaddress
%option in \documentclass). \noaffiliation is required (may also be
%used with the \author command).
%\collaboration can be followed by \email, \homepage, \thanks as well.
%\collaboration{}
%\noaffiliation

\date{\today}

\begin{abstract}
All cells derive nutrition by absorbing some
 chemical and energy resources from the environment;
these resources are used by the cells to reproduce the chemicals within them,
which in turn leads to an increase in their volume.
In this study we introduce a protocell model exhibiting
 catalytic reaction dynamics, energy metabolism, and cell growth.
Results of extensive simulations of this model show the existence of
 four phases with regard to
 the rates of both the influx of resources and cell growth.
 These phases include an active phase with high influx and high growth rates,
 an inefficient phase with high influx but low growth rates,
 a quasi-static phase with low influx and low growth rates,
 and a death phase with negative growth rate.
A mean field model well explains
 the transition among these phases as bifurcations.
The statistical distribution of the active phase is characterized
 by a power law, and that of the inefficient phase is characterized
 by a nearly equilibrium distribution.
We also discuss the relevance of the results of this study
 to distinct states in the existing cells.
\end{abstract}

% insert suggested PACS numbers in braces on next line
% insert suggested keywords - APS authors don't need to do this
%\keywords{}

%\maketitle must follow title, authors, abstract, \pacs, and \keywords
\maketitle

% body of paper here - Use proper section commands
% References should be done using the \cite, \ref, and \label commands
\section{\label{sec:level1}Introduction}
To unveil the minimal requirement of life, we need to first understand
 how an ensemble of a variety of catalytic reactions generates a stable cell
 that grows to reproduce itself.
A cell in any form---right from its prototype form at its inception
 to its currently evolved form---receives an influx of some nutrient chemicals;
the cell then effectively uses the energy obtained from the decomposition
 of these nutrient chemicals to develop useful catalysts and membranes.
It should be noted that the chemicals and energy obtained from the nutrients
 through catalytic reactions are used for cell growth and cell replication.

Given the fact that cell growth requires a continuous flow of energy and nutrients,
 it can be said that cell growth will not be sustained
 if the cell is in thermal equilibrium. Indeed, the relevance of non-equilibrium open systems to life processes has been discussed
 over several decades\cite{book:schrodinger,book:prigogine}.
On the other hand, in a Carnot cycle, the maximal thermodynamic efficiency is achieved under an adiabatic condition,
 where a weak influx flow is generated under a condition close to the equilibrium condition.
In this case, a cellular state far from equilibrium may not be appropriate for efficient thermodynamic
processes. 
Therefore, it is pertinent to determine the condition under which highly efficient cellular
 growth will be 
achieved\cite{jbr1995mr}.

This problem regarding efficiency in cell-volume  growth is a general problem so that it is
essential to understand the primitive form of a cell, or a protocell. 
In this case, one needs to understand a universal property of a growing cell
following catalytic chemical reaction processes upon external nutrients.
On the other hand,  this issue is also important to the present cells.
In this case, it is likely that the rate of conversion of chemical resources to increase the cell volume
 depends on the current cellular state, or one may distinguish distinct cellular states
 on the basis of the efficiency of conversion.

With regard to such cellular states, it is a known fact that a cell undergoes several ``phases''
 depending on culture conditions,
 where each phase is characterized by a distinct growth rate\cite{jm1949arm,nqb2004sci,ik2004jb}.
The appearance of these phases depends on the nutrient condition; moreover,
 each phase represents a distinct cellular state whose chemical composition differs
 from that of any other state.
For example, in the case of bacteria,
a high growth rate is achieved in the log phase, which is characterized by
 exponential cell growth under a nutrient-rich condition;
having said that, bacteria also undergo some secondary phases such as a stationary phase
 characterized by suppressed growth and a dormant phase characterized by suppressed metabolism.
These observations imply the existence of qualitatively distinguishable cellular states.
With this background, is it now possible to understand the existence of such distinct phases,
characterized by distinct growth and influx rates, in a cell?  The purpose of the present study is
to introduce a simple ``protocell'' model having distinct cellular states and
analyze efficiency for cellular growth depending on each state.

To investigate the origin of multiple cellular states
 and underlying energetics theoretically,
 a cell model including metabolism and growth is needed.
There are extensive studies conducted on protocell models that are based on catalytic reactions
\cite{book:hypercycle,tg1975bs,rjb1991al2,ds1998oleb,sj2001pns,cf2003prl,cf2005bp,kk2005acp,cf2006pre,tr2007ptb,tc2008jtb} to determine possible conditions that facilitate the reproduction of cells.
Previously, Furusawa and one of the authors (KK) studied a protocell model
 consisting of a large number of chemicals that mutually catalyze each other,
 in order to understand the manner in which a cellular state is realized
\cite{cf2003prl,cf2005bp,kk2005acp,cf2006pre}.
Random catalytic networks were used to identify the condition for steady cell growth,
 where we found a universal statistical law on the abundance distribution of chemicals.  Interestingly,
in spite of the simplicity of the model with just catalytic reaction networks, this statistical law 
was also confirmed over most of the existing cells \cite{cf2003prl}.

However, in the model, energetics was not discussed.
It was assumed that the energy required for catalytic reactions was supplied automatically, and also
that the growth rate of the cell  was equal to the inflow rate of the resources.
Furthermore, multiple distinct cellular states were not observed
 with regard to this model.
Therefore, the growth rate was uniquely determined by
 the nutrition condition, and the efficiency of cell growth depending on  cellular states could not
be studied at all.

To study the cell-state dependence of growth efficiency, we extend here
the above-mentioned protocell model to include energy metabolism.
For this purpose, we first need to consider reversible reactions,
 with the reaction rate constants given by the chemical potentials of molecules
 to be consistent with thermodynamics \cite{me2007bpj,aa2009pre}.
This consideration is in contrast to that pertaining to most models
 of catalytic reaction networks
\cite{book:hypercycle,cf2003prl,cf2005bp,kk2005acp,cf2006pre,sak1986jtb,pfs1993phd,sj2001pns,rh2005pre},
 where only irreversible(uni-directional) reactions are considered.
Second, we introduce specific chemicals for ``energy currency molecules'' such as adenosine triphosphate (ATP),
 to derive energy from the reactions or to supply energy to the reactions.
This energy currency component is assumed to be used for cell growth and
 is synthesized by other catalytic reactions.
In fact, the cell growth usually requires some high-energy chemicals
, the synthesis of which requires energy currency molecules.
For example, the synthesis of DNA requires ATP consumption.

In the case of our model, first we consider that there is an influx of nutrient resources.
These nutrients are used in the synthesis of a variety of catalysts
 through reactions given by catalytic reaction networks.
An energy currency molecule (e.g., ATP) is synthesized through such reactions,
 while the cell volume increases by consuming the energy generated by this molecule.
Further, the metabolic efficiency of this model, i.e., the rate of cell growth
 for consumption of nutrients, depends on the internal cellular state,
 i.e., the concentrations of the catalysts.
Hence we can study the relationship between the influx rate and the cell growth rate,
 depending on the cellular state,
 and examine the possibility of existence of distinct cellular states with unique
 metabolic efficiencies and cell growth rates.

The present paper is organized as follows.  In the next section we introduce a model
consisting of  a catalytic reaction network with  energetics, and cellular growth.
In Sect. III, direct simulation of the model and the mean-field analysis are
presented, where we report the existence of different phases with
distinct cellular growth speed and efficiency, as well as different statistical
characteristics of chemical abundances.  Section IV is devoted to
summary and discussion.

% Put \label in argument of \section for cross-referencing
%\section{\label{}}

\section{Model}
\begin{figure}[H]
  \begin{center}
    \includegraphics[height=5.4truecm]{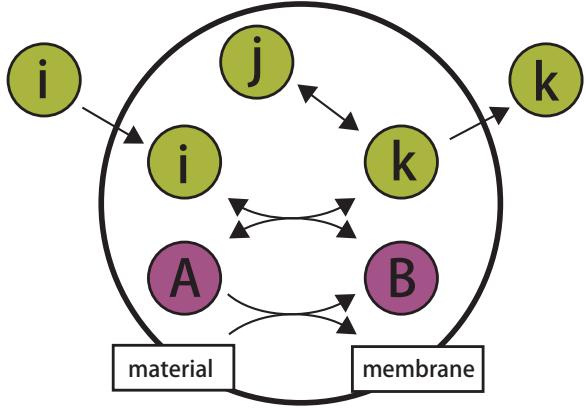}
  \end{center}
  \caption{
    Schematic representation of our ``protocell'' model.
    It consists of an encapsulated chemical reaction network within a membrane.
    Green circles indicate catalytic components
    and $i,j,k$ are the component indices;
    purple circles indicate energy currency molecules.
    Arrows across the membrane represent material exchange by diffusion through the membrane,
    whereas arrows within the membrane represent chemical reactions.
    Conversion from energy currency component $A$ to $B$ is used for
    the synthesis of membrane molecules resulting in an increase in the cell volume.
  }
  \label{fig:model}
\end{figure}

Here, we consider a very primitive ``minimal'' protocell, compartmentalized by a membrane,
 within which encapsulated catalytic reactions progress
 (see Fig.\ref{fig:model} for schematic representation).
This catalytic reaction is necessary for the synthesis of all the chemicals within the cell,
 and also brings about cell growth.
For this synthesis, some supply of matter and energy is generally required.
For the former, we consider some nutrient chemical species, from which catalysts
 are synthesized.
For the latter, a common molecular species that works as an ``energy currency'', say ATP, is assumed;
 it is required for many endothermal reactions to progress.
There are influxes and effluxes of these and other chemicals.
As the number of molecules increases, the cell volume increases (as a result of synthesis of membrane molecules)
 and, accordingly, the concentrations of all chemicals within the cell decreases.

For system variables, we consider the following:
 (i) A set of concentrations of catalytic components whose concentrations are
 given by $(x_0,x_1,...x_{N-1})$.
These chemicals are synthesized by catalytic reaction(s),
 and other metabolites are assumed to exist constantly, hence
 they appear as constant parameters in the reaction.
 (ii) An energy currency molecule. With ATP and ADP in mind,
 we assume that the conversion of A to B supplies energy (A corresponds to ATP, and B to ADP).
 The concentration variables of these are $(y_A,y_B)$.
 (iii) The volume of the protocell is set at $V$, and increases or decreases with time.

{\sl Catalytic reaction coupled to the conversion of energy currency molecules}:
 We consider a reaction process from $i$ to $j$ catalyzed by $l$,
 coupled with the conversion of energy currency molecules.
 This reaction is given by
\begin{equation}
  i + l + A(B) \leftrightarrow j + l + B(A),\label{eq:catalytic_reaction}
\end{equation}
where $i,j,l$ are reactant, product, and catalyst, respectively.
By denoting the standard chemical potentials for the catalytic components and energy currency components
 as ($\mu_i^0, \mu_j^0, \mu_A^0, \mu_B^0$),
 the reaction rate constant satisfying the detailed balance condition is determined as
\begin{equation}
  k_{ij} = A_{ij}\min[1,e^{-\beta (\mu_j^0 + \mu_{B(A)}^0 - \mu_i^0 - \mu_{A(B)}^0)}],\label{eq:rate_constant}
\end{equation}
where $\beta$ is the inverse temperature and $A_{ij} = A_{ji}$\cite{aa2009pre}.
The reaction rate for $i$ to $j$ is then represented as $k_{ij}x_i x_l y_{A(B)}$.
Note that the reaction progresses to reduce the free energy of the associated chemicals, not the energy.
From this point of view,
 the assumptions of the mass-action law and the rate constants
 correspond to taking the form of the chemical potential as $\mu_i \equiv \mu_i^0 + (1/\beta) \log x_i$.
In fact, the net reaction flux from $i$ to $j$, i.e., $x_l(k_{ij}x_iy_{A(B)} - k_{ji}x_jy_{B(A)})$,  always reduces
 the free energy per unit volume $g = x_i\mu_i + x_j\mu_j + y_A\mu_A + y_B\mu_B$.

We consider a variety of reactions.
The whole set of reactions constitutes a network given by the tensors
 $C_A(i,j,l)$ and $C_B(i,j,l)$, which represent the structure of a catalytic reaction network, defined by
\begin{equation}
  C_{A(B)}(i,j,l) =
  \begin{cases}
    1 & ({\rm If\ reaction}\ i + l + A(B) \to j + l + B(A){\rm\ exists})\\
    0 & ({\rm otherwise})
  \end{cases}
\end{equation}

The nutrient chemical is assumed to diffuse through the membrane,

 and its external concentration is set at $X_0$, hence
 the efflux rate is given by $D_0(X_0 - x_0)$,
 with the diffusion coefficient $D_0$.
For simplicity, other components except the chemical $N-1$ are assumed to be impermeable, i.e, $D_i = 0$.
The external concentration $X_{N-1}$ is set much smaller than $X_0$,
 hence the cell takes $0$ as nutrient inflow, and diffuses out $N-1$ as waste.
However, this condition is not essential and can be relaxed.
Several other $D_i$s could be set at non-zero,
 and the result would be qualitatively unchanged.

Cell growth progresses as a result of membrane increases(and the influx of water in proportion).
Instead of explicitly introducing a membrane molecule, here we simply note
 that membrane growth needs some energy currency consumption.
Assuming that the precursor molecules for the membrane are
 abundant in the medium, we consider a phenomenological ``growth reaction'' that progresses
 accompanied by the reaction $A \to B$.
We assume that the rate of the growth reaction also satisfies the detailed balance condition.
Then, defining $P$ and $\mu_{P}^0$ as the concentration
 and standard chemical potential, respectively, of the membrane precursor,
 and $\mu_M$ as the (effective) chemical potential of the membrane,
we write the rate of the growth reaction as
\begin{equation}
  k(P y_A e^{\beta (\mu_P^0 + \mu_A^0)} - y_B e^{\beta (\mu_M + \mu_B^0)})
\end{equation}
 where $k$ is a rate constant.
In addition, we assume that the concentration of the precursor is fixed.
Hence, the volume growth is simply given by
\begin{equation}
  \frac{1}{V}\frac{dV}{dt} = \gamma k_g(y_A - y_Be^{\beta \epsilon})\label{eq:growth_reaction}
\end{equation}
 where $\epsilon$ $(= \mu_M + \mu_B^0 - \mu_P^0 -(1/\beta)\log P - \mu_A^0)$
 is the energy required for the process,
 $k_g$ $(= kPe^{\beta (\mu_P^0 + \mu_A^0)})$ is its rate constant,
 and $\gamma$ is the conversion rate from the membrane molecules to the cell volume.

Here, $P,\mu_P^0,\mu_M,\mu_A^0,\mu_B^0,k$
 are direct independent parameters controlling the reactions,
 and $k_g,\epsilon$ are functions of these parameters.
However, the behaviors of the model as dynamical systems are well represented by
 choosing $k_g, \epsilon$ (and $X_0$) as relevant control parameters,
 hence that we use these as the relevant parameters in the present model.

As a result of the growth in cell volume $V$, given by eq. (\ref{eq:growth_reaction}),
 all the chemicals are diluted accordingly.
This dilution leads to a decrease in the concentrations of all the chemicals in proportion to themselves,
 i.e., $x_i\frac{1}{V}\frac{dV}{dt}$.
As a result of this dilution, it is expected that the total concentration of the energy currency
 components ($y_A + y_B$) would decreases, and, without a supply, the system would cease to grow.
In reality, such molecules would also be diffused from the outside, say phosphorate molecules,
 the material components for ATP and ADP.
Hence, the energy currency molecules are assumed to flow in
 with increasing the cell volume so that $y_A + y_B$ is constant.
Consequently, the two concentrations are not independent.
We therefore only need to consider the time evolution of $y_A$.

Taken together, the time evolution of the concentrations of the catalysts $x_i$s
 and energy currency $y_A$ are given by
\begin{eqnarray}
\nonumber  \frac{dx_i}{dt} &=& \sum_{(n,m) = (A,B),(B,A)}
\left( \sum_{i,j,l} C_n(j,i,l)x_l(k_{ji}x_jy_{n} - k_{ij}x_iy_{m})\right)\\
  &+& D_i(X_i - x_i) - x_i\frac{1}{V}\frac{dV}{dt}\label{eq:x}
\end{eqnarray}
\vspace{0.2truecm}
\begin{eqnarray}
\nonumber  \frac{dy_A}{dt} &=& -k_g(y_A - y_Be^{\beta \epsilon})
 + \sum_{i,j,l} C_B(i,j,l)x_l(k_{ij}x_iy_B - k_{ji}x_jy_A).\label{eq:y}
\end{eqnarray}
\vspace{0.2truecm}

First, all of the coefficients of the chemical reaction rate ($A_{ij}$s) are set at 1, for simplicity.
Secondly, the standard chemical potentials of the catalysts ($\mu_i^0$s)
 are sampled from a uniform distribution from 0 to 1.
Thirdly, if the component 0 is a nutrient,
 the energy currency A is assumed to be synthesized when the component 0 is degraded,
 by appropriately setting the directions between the catalytic reactions and the conversion of
 the energy currency component.
For all other reactions, the coupling directions have been determined randomly.

Finally, we make a cautious comment on the effect of the influx of the energy currency molecules
 on the intracellular energy currency composition.
The compositions of the energy currency molecules $A$ and $B$ outside the cell
 are expected to be in chemical equilibrium.
Then, when the energy currency molecules in such a composition
 flow in with cell growth,
 the compositions of the intracellular energy currency components ($y_A, y_B$)
 will be equilibrated.
We can treat this effect by $(y_A - y_A^{eq})\frac{1}{V}\frac{dV}{dt}$
 onto the time evolution of $y_A$, where
 $y_A^{eq}$ is the concentration of $y_A$ at equilibrium,
 and we carried out simulations for such a model.
However, we confirmed that the results discussed in the present paper
 are qualitatively the same, independent of this effect;
hence we neglect this effect for simplicity.

\section{Results}
\subsection{\label{sec:level2}Emergence of states with different metabolic efficiencies}

We simulated the present model using several parameter values and
 taking a variety of catalytic reaction networks.
Each network was generated randomly with a given total reaction number $G$
 under the constraint that it is percolated.
From these simulations, we found common behaviors in the present protocell model,
 as long as the reaction path number was sufficient.
Indeed, as long as the average path number per chemical, i.e.,
 $K = G/N$, satisfies  $K > K^* \sim 8$, the common behavior is observed,
 independent of the network.
If $K$ is smaller than that, the behavior often depends on the network, which we
do not discuss here.

For most of these cases, the chemical concentrations and growth rates
 converge to a fixed point after some relaxation time.
If the growth rate is positive, a fixed point represents a cellular state with steady growth.
If it is negative, the cell shrinks, and will ultimately collapse.
We are interested in the former case here.
Since the approach to fixed points is quite common, we focus on the properties of such fixed points,
 without going into details of the transient relaxation process.
For most parameter values, the stable fixed point is unique,
 but at some parameter values, there are two stable fixed points, leading to bistability
 in the cellular state;this is discussed below.

In cell state characterization, the growth rate $\lambda \equiv \frac{1}{V}\frac{dV}{dt}$
 and the influx $J_{in} \equiv D_0(X_0 - x_0)$ are important order parameters for characterizing the protocell;
 the ``metabolic efficiency'' $\eta$, defined as $\eta \equiv \lambda /J_{in}$,
 is another basic order parameter.

With regard to the internal properties of a protocell, the density $m$ is defined as the total
 concentration of catalysts, as $m \equiv \sum_i x_i$.
The density is an important order parameter and is related to influx, efflux, and growth rate.
In a steady state of the cell, there has to be a balance between material influx and dilution
 by cell volume growth so that the density is constant.
Hence, we get
\begin{equation}
  \lambda m = J_{in} - J_{out}
\end{equation}
which gives the relationship among the growth rate, density, influx, and efflux.
Then, the efficiency $\eta$ is represented by
$\eta = (1 - J_{out}/J_{in})/m$.
The three order parameters ($\lambda,J_{in},m$) are
 sufficient to specify a cellular state macroscopically.

\begin{figure}[H]
  \begin{center}
    \subfigure[]{\includegraphics[height=3.5truecm, width=4.7truecm]{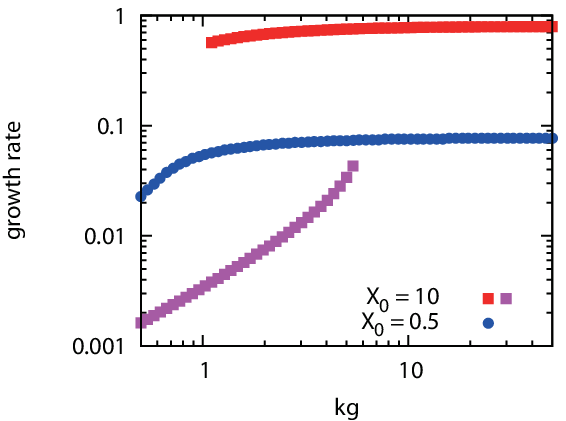}}
    \subfigure[]{\includegraphics[height=3.2truecm, width=4.2truecm]{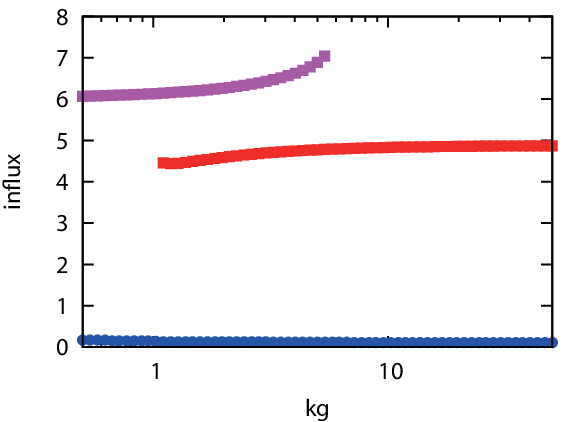}}
    \subfigure[]{\includegraphics[height=3.5truecm, width=4.6truecm]{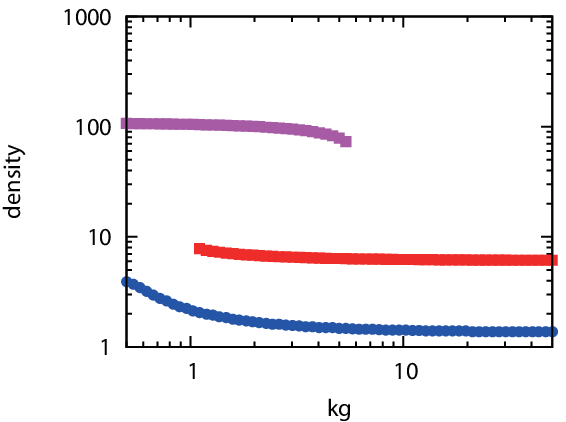}}
  \end{center}
  \caption{
Dependences of the steady state properties on growth reaction constant $k_g$ for $X_0 = 0.5$ (blue)
 and $X_0 = 10$ (red$/$purple),
 using a catalytic network with $N = 24, K = 10$.
The parameters were set as
 $X_{N-1} = 0.01$, $D = 1.0$, $\gamma = 1.0$, $\beta = 1.0$, $\epsilon = 0.0001$,
 $\mu_A^0 = \mu_B^0 = 1.0$, and $y_A + y_B = 1.0$.
 (a)Growth rate:$\lambda = \frac{1}{V}\frac{dV}{dt}$;
 (b)influx:$J_{in} = D_0(X_0 - x_0)$;
 and (c)density:$m = \sum_i x_i$.
  }
  \label{fig:fix_point}
\end{figure}

Fig. \ref{fig:fix_point} shows the dependences of the growth rate $\lambda$, influx $J_{in}$, and density $m$
 on the growth reaction rate constant $k_g$
 for a certain catalytic reaction network with $N = 24, K = 10$.
(Indeed, this behavior is commonly observed, independent of the networks.)
When $X_0$ is large ($X_0 = 10$), it is found that there are two branches of
 fixed points with different growth rates, as shown in Fig. \ref{fig:fix_point}(a).
In this case, these two fixed points are bistable when $k_g$ takes an intermediate value.
The transition between the two states with the increase or decrease in $k_g$ occurs with hysteresis.

These two states with higher/lower growth rates correspond to lower/higher influx
 and density, respectively, as shown in Figs. \ref{fig:fix_point}(b) and \ref{fig:fix_point}(c).
Here, however, the dependence of the influx on $k_g$ is much weaker, and
 there is not a large gap between the influx values of the two states.
Thus, the difference between the growth rates of the two states directly represents
 the difference in metabolic efficiency $\eta$.
This significant difference in the efficiency reflects the internal
 state of the cell, as discussed later.

Fig. \ref{fig:fix_point} also shows that only one
 state with the intermediate growth rate exists when the external nutrient
 concentration is small ($X_0 = 0.5$).
This state has a very low influx value, and accordingly, a high efficiency
 and is therefore distinguishable from the above two states.
When $X_0$ decreases, we observed that the efficiency increases, which is
 a common behavior, independent of the values of $k_g$.

Besides the states shown in Fig. \ref{fig:fix_point}, another
 state with a negative growth rate appears when $k_g$ is large enough.
In this case cells shrink, and hence that the state is regarded as a ``death'' state.
We note that the ``death'' state coexists with another growing state
if $X_0$ is not small, but, for small $X_0$ values ( say smaller than 0.1),
 it is the only attractor.

\begin{figure}[H]
  \begin{center}
  \subfigure[]{\includegraphics[height=4.8truecm]{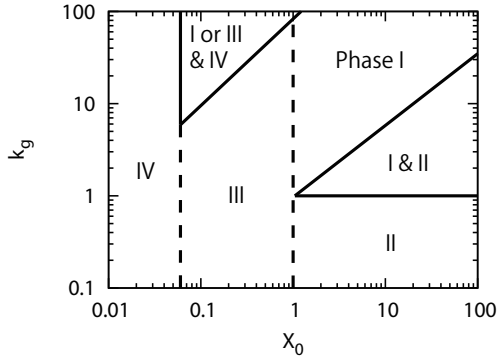}}
  \end{center}
  \caption{
    Phase diagram of the model on the $X_0-k_g$ plane.
    Each phase (I,II,III,IV) is defined as in Table \ref{table:phase}.
    Solid lines represent discontinuous transitions between phases;
    broken lines represent continuous transitions.
  }
  \label{fig:pd_model}
\end{figure}

Summarizing these distinct states, we have plotted the phase diagram
 against $k_g$ and $X_0$ as shown in Fig. \ref{fig:pd_model}.
Each ``phase'' here is defined and characterized as in Table \ref{table:phase}.
In particular,
 as expected from the plot in Fig. \ref{fig:fix_point},
 the range of $k_g$ values in which phases I and II coexist
 narrows continuously as $X_0$ decreases, and finally disappears.

\begin{table}[H]
\begin{center}
\begin{tabular}{|l|l|l|l|l|}
\hline
Phase&$\lambda$&$J_{in}$&$\eta$&Distribution of abundances\\ \hline
I: Active phase& High& High&Medium&Power law\\ \hline
II: Inefficient phase& Low&High& Low&Close to equilibrium\\ \hline
III: Quasi-static phase& Low& Low& High&Mixed\\ \hline
IV: Death phase&Negative& & &\\
\hline
\end{tabular}
\end{center}
\caption{
  Definition of phases and characterization by order parameters,
 and the statistical distribution of abundances of chemicals to be discussed in Sec. \ref{subsec:rank}.}
\label{table:phase}
\end{table}
\subsection{Mean field analysis}

The existence of each phase and the structures of the transitions between phases
 are explained by carrying out a simple mean field approximation for our model.
To construct the mean field equations of our model,
 we consider the average concentration $x$ of chemicals over all chemical
 components ($i = 0,1,...,N-1$), i.e., $x=m/N$, and replace the rate equation by this concentration.
To get this mean field approximation, the catalytic network is approximated by a fully
 connected network with weak paths
 so that it has the the same average path density $\rho$ in the original model (= $K/N$).
Also, the standard chemical potentials $\mu_i^0$ of all the chemicals are set to be identical
 (to their mean value), and
 the standard chemical potentials of the energy currency components $\mu_A^0,\mu_B^0$ are the same.
In addition, by assuming that diffusion through the membrane is fast enough,
 the influent component concentration $x_0$ and effluent component concentration $x_{N-1}$
 are regarded to be equilibrated with their external values,
 and they are replaced by $X_0$ and $X_{N-1}$ respectively.

Combining these approximations, we obtain the following temporal evolution equations,
 which consist of the concentration variables of the catalyst $x$
 and the energy currency molecule $y_A$, as
\begin{equation}
  \frac{dx}{dt} = \rho x(X_0y_B - xy_A) - \frac{1}{2}\rho x(y_A + y_B)(x - X_{N-1})
 - xk_g \gamma (y_A - y_Be^{\beta \epsilon})\label{eq:mfe_x}
\end{equation}
\begin{eqnarray}
\nonumber  \frac{dy_A}{dt} &=& \rho (N-2)x(X_0y_B - xy_A) - \frac{1}{2}\rho (N-2)x(y_A - y_B)(x - X_{N-1})\\
&-& k_g(y_A - y_Be^{\beta \epsilon}) - \rho (N-2)^2x^2(y_A - y_B).\label{eq:mfe_y}
\end{eqnarray}

\begin{figure}[H]
  \begin{center}
  \subfigure[]{\includegraphics[height=4.8truecm]{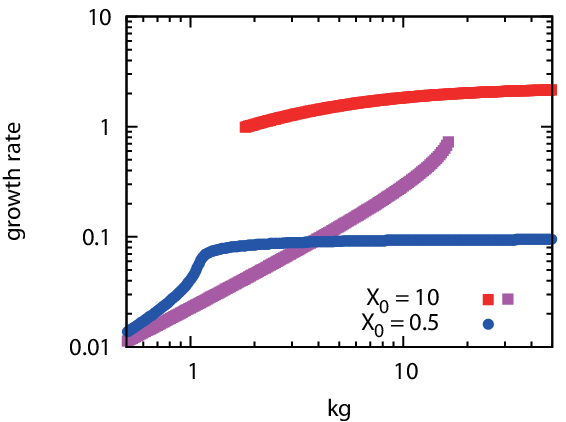}}
  \subfigure[]{\includegraphics[height=4.8truecm]{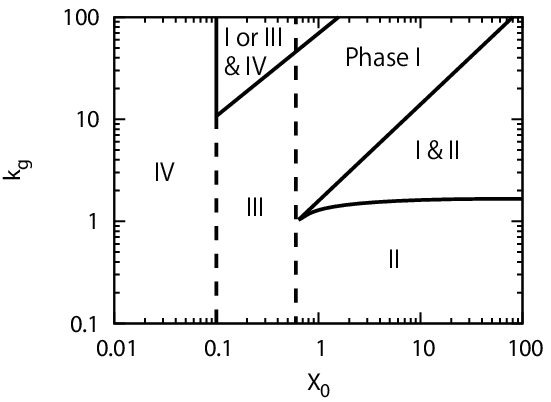}}
  \end{center}
  \caption{
    Behaviors of the mean field equations (eqs. (\ref{eq:mfe_x}) and (\ref{eq:mfe_y})).
    The other parameters are set to be identical with those adopted in Fig. \ref{fig:fix_point}.
    (a)
    Dependence of growth rate on $k_g$, obtained from the steady states of the mean field equations,
    for $X_0 = 0.5, 10$.
    Colors represent different $X_0$ values.
    (b)
    Phase diagram from mean field equations on the $X_0-k_g$ plane.
    Each phase is defined in the text as given in the original model.
    Solid lines represent
    discontinuous transitions between the phases
    with saddle-node bifurcations, and
    broken lines represent continuous transitions distinguished by
    changes in the growth rate and influx without bifurcations.
  }
  \label{fig:mfe}
\end{figure}

As with the direct numerical results, the existence of two branches of fixed points
 is confirmed for large $X_0$, with hysteresis as a change of the growth reaction rate constant $k_g$.
In Fig. \ref{fig:mfe}(a), the growth rates are plotted against $k_g$ for $X_0 = 0.5$ and $10$,
 which are obtained from the mean field equations, eqs. (\ref{eq:mfe_x}) and (\ref{eq:mfe_y}).
Comparison with the results in Fig. \ref{fig:fix_point}(a) indicates that the obtained dependence of
 the two fixed point values on the parameter $k_g$,
 and the existence of hysteresis for large $X_0$,
 agree well qualitatively with the results from the original model.

In addition, the death phase, IV, with a negative growth rate, is also obtained from the mean field equations.
Again, for a medium value of $X_0$, this fixed point coexists with that for phase I or III.
In contrast, with decreasing $k_g$, the region of coexistence narrows and
 finally disappears.

By summing all the solutions, a phase diagram is obtained from the mean field equations,
 as shown in Fig. \ref{fig:mfe}(b);the diagram agrees rather well with that
in the original model(Fig. \ref{fig:pd_model}).

From the mean field equations, one can see that the mean catalyst
 concentration $x$ (or the density $m = Nx$) has a negative effect on the growth rate $\lambda$
; this introduces bistability in phases I and II.
The third and fourth terms in the right-hand side of eq. (\ref{eq:mfe_y}) indicate
 the consumption of $A$ by the growth reaction, and the degradation
 of $A$ by catalytic reactions, respectively.
The dependence of each term on catalyst concentration $x$ shows that
 the degradation of $A$ increases faster than consumption,
 so that the allocation of energy currency to the growth processes decreases with increasing $x$.
Hence, for large $x$, the abundance of $A$ and the growth rate are suppressed.
Consequently, the dilution effect, represented by the third term in the right-hand side of eq. (\ref{eq:mfe_x}),
 does not work effectively, and $x$ is sustained to take a large value.
On the other hand, when $x$ is initially small, it continues to take a small value,
 as it is suppressed by dilution.
As a result, two states with different growth rates coexist.
Noting that $m=xN$ in the mean field approximation and
 replacing $x$ by the density $m$,
 the above argument is also applied to the original model.
To sum up, there is mutual inhibition between the growth rate and catalyst density.
This mutual inhibition leads to bistability.
Note, however, that for this mechanism to work, the density $m$ should be sufficiently high,
 so that bistability is not present for small $X_0$.
In this case, a single phase, III, exists against the change in $k_g$.

Since the mean field equation involves just two variables, the transitions among each phase
 are analyzed as bifurcations of fixed points.
From the standard analysis of the flow of a two-dimensional state space and
 the linear stability of the fixed points,
 it is straightforward to show that there are saddle-node bifurcations at the transitions
 from phase I to II and II to I, with $k_g$ as a bifurcation parameter.
These two bifurcation points correspond to the boundaries between the
 phase I and the bistable region, and the phase II and the bistable region.
As $X_0$ decreases, these bifurcation points approach and finally collide and disappear.
This disappearance corresponds to the transition to phase III.

Transition to the death phase, IV, has the same bifurcation structure.
There are saddle-node bifurcations at the transition
 between the death phase IV and the growing phases with $X_0$ as a bifurcation parameter.
These bifurcation points approach and finally collide with each other and disappear
 as $k_g$ decreases.

\subsection{Statistical natures of internal states}\label{subsec:rank}

So far, we have shown the distinct phases characterized by ``macroscopic'' variables,
 such as the growth rate, influx, and total density of chemicals, which
 are well represented by the mean field equations.
In a direct simulation involving many chemicals, however, there are `microscopic'
 variables such as the catalyst concentrations ($x_0,x_1,...x_{N-1}$).
In our model, there is no specific meaning to each chemical,
 hence the distribution of abundances is a relevant characteristic of a cellular state.
For this statistical property of the concentrations, we take a
 rank-ordered concentration distribution,
 i.e., we plot the abundances in order of magnitude, following
 earlier studies\cite{cf2003prl}.
As a result, it is revealed that each phase
 has a quite different statistical property as discussed below.

\begin{figure}[H]
  \begin{center}
    \subfigure[]{\includegraphics[height=4.4truecm]{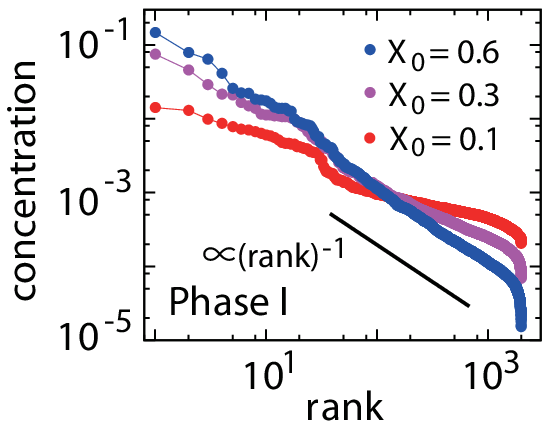}}
    \subfigure[]{\includegraphics[height=4.6truecm]{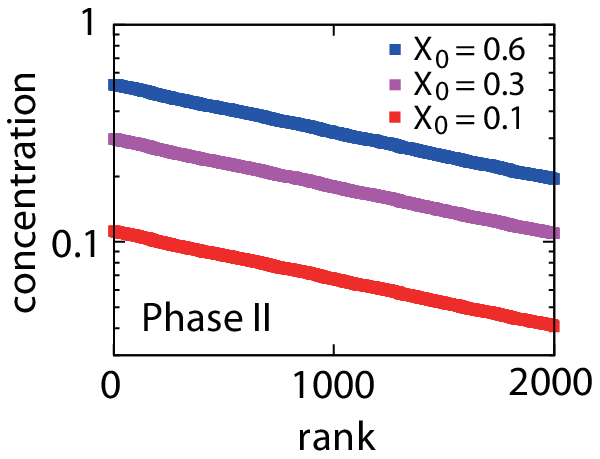}}
    \subfigure[]{\includegraphics[height=4.6truecm]{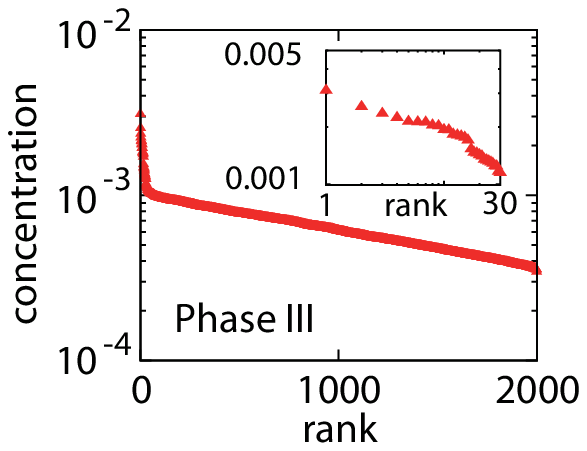}}
    \subfigure[]{\includegraphics[height=4.4truecm]{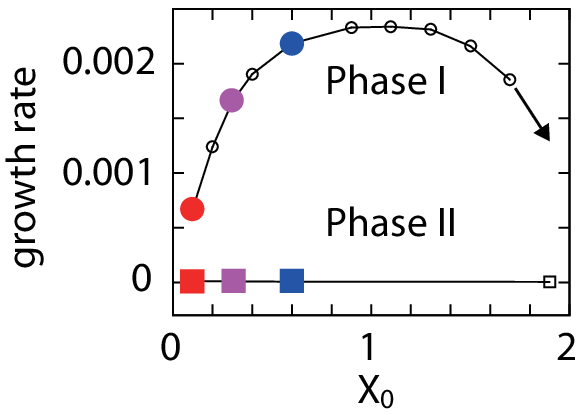}}
    \subfigure[]{\includegraphics[height=4.4truecm]{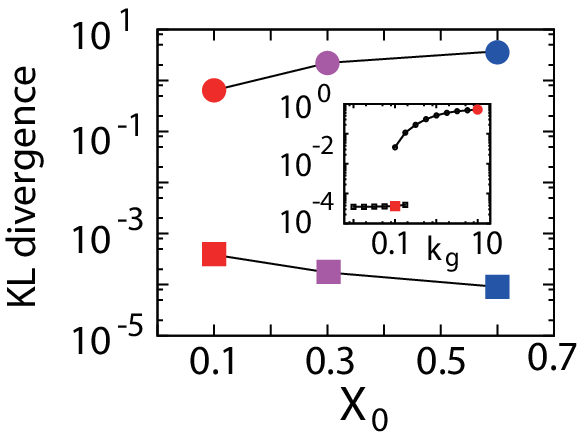}}
  \end{center}
  \vspace{-0.5truecm}
  \caption{
Statistical nature of the model.
A network with $N=2000$ and $K = 20$ was used.
 (a)Rank-ordered concentration distributions in phase I for $X_0 = 0.1,0.3,0.6$.
Colors represent different values of $X_0$.
Here $k_g = 10.0$, and the other parameters are same as those in Fig. \ref{fig:fix_point}.
(b)Distributions in phase II.
The network and parameters are the same as those in (a), except $k_g = 0.1$.
(c)A distribution in phase III.
The network and parameters are the same as those in (a), except for $X_0 = 0.02, k_g = 5.0$.
The inset shows a log-log plot of the distribution up to rank 30.
(d)Dependences of the growth rate on $X_0$ in phases I and II.
Colors and shapes of points represent data corresponding to the plots
 in (a) and (b) sharing the same color and shape.
(e)Dependences of Kullback-Leibler divergence from the distributions to
 the equilibrium compositions on $X_0$.
The inset shows the dependencies on $k_g$.
  }
  \label{fig:rank}
\end{figure}

Figs. \ref{fig:rank}(a), (b), and (c) show rank-ordered concentration distributions
 in phases I, II, and III, respectively.
The ordinate indicates the concentration of catalysts and the abscissa
 shows the rank determined by the concentration.
For phase I, we observed that the slope of the rank-ordered concentration distribution
 increases with increments of $X_0$, and reaches a power-law distribution
 with exponent -1.
However, we note that this power-law distribution with an exponent -1,
 in other words,
 Zipf's law, can be achieved only when
 the catalytic reaction network is sufficiently sparse
 (approximately $K < 40$ for $N=2000$).
For a dense network,
 although the slope of the distribution increases with $X_0$,
 the distribution remains curved in the log-log plot,
 and thus does not fall on the power law.
The change in the slope is accompanied by a change in growth rate,
 which, for reference, is plotted in Fig. \ref{fig:rank}(d)
 with increasing $X_0$ for phase I.
Each point in the figure corresponds to each distribution in
 Figs. \ref{fig:rank}(a) and \ref{fig:rank}(b) of the same color and shape.
With increasing the growth rate,
 the slope of the distribution also increases and approaches
 the power law with exponent -1, near the optimal growth rate.
With further increases in $X_0$, however,
 the growth rate decreases until it reaches zero.
In this region before the arrest of growth,
 the concentrations approximately obey a power-law distribution
 with exponents -1 or slightly higher(data not shown).
In this region, however, the steady state is often no longer stable,
 and the growth rate shows oscillations with time.
The plot with the downward arrow in Fig. \ref{fig:rank}(d) denotes
 a regime with such oscillations.

Fig. \ref{fig:rank}(b) shows rank-ordered distributions
 in phase II, for several values of $X_0$ by a semi-log plot.
In contrast to phase I, the plot is linear on a semi-log scale,
 and the slope is independent of the values of $X_0$.
In other words, the distribution is exponential,
 which, as shown below, comes from the fact that
 the concentrations have near equilibrium compositions in these states,
 defined by the standard chemical potentials of the catalysts as $x_i^{eq} \propto \exp (-\beta \mu_i^0)$.
Because of the uniform distribution of the potentials $\mu_i^0$,
 the rank-ordered equilibrium distributions take the form of
 exponential distributions.

To confirm that the distributions in phase II are actually
 close to the equilibrium distribution,
 we measure the distance from equilibrium
 by calculating the Kullback-Leibler divergence from
 the equilibrium distributions of the concentrations
 ($ = \sum_i p_i\log (p_i/p_i^{eq})$, where $p_i^{(eq)} \equiv x_i^{(eq)}/\sum_i x_i^{(eq)}$)\cite{es2008jtb}
 as plotted in Fig. \ref{fig:rank}(e), against the change in $X_0$
; the change of the divergence against $k_g$ is shown in the inset of Fig. \ref{fig:rank}(e),
 ranging between phase I and II.
It is shown that the divergence in phase II is quite low compared
 to that in phase I and decreases with increments of $X_0$,
 to approach equilibrium.
Although the emergence of exponential distributions itself
 depends on the assumption of uniform distribution of $\mu_i^0$s,
 the decreases in the divergence are independent of assumptions
 on the distribution of $\mu_i^0$s.
Thus, phase II can be characterized by the distribution
 close to the thermal equilibrium.
As shown in the inset of Fig. \ref{fig:rank}(e),
 this branch connected continuously from the equilibrium
 shows a discontinuous jump to phase I,
 where the composition of the concentration distribution
 is far from that at thermal equilibrium.

Thus, phases I and II are characterized by distributions with
 power-law distributions and close to equilibrium, respectively.
On the other hand, the states in phase III are mixed, as shown in Fig. \ref{fig:rank}(c).
For chemicals with smaller abundances the rank-abundance plot follows exponential decrease
 and is near equilibrium, as in phase II.
The concentrations of abundant chemicals, however, deviate distinctly from the
 equilibrium distribution. The rank-abundance plot for abundant chemicals
 follows a power law, as in phase I, while the fraction of such chemicals that deviates from
 equilibrium distribution increases with $k_g$.
This partial deviation from equilibrium distribution characterizes phase III.

\section{Summary and Discussion}
We have studied a simple ``protocell'' model, in which
 chemicals are encapsulated within a membrane.
The protocell synthesizes these chemicals by using the nutrients it derives from
 the environment;the protocell also utilizes the energy obtained from
 intracellular reactions to increase its volume.
These reactions are catalyzed by other chemicals that are also synthesized within the cell,
 and thus, the total set of reactions constitutes a catalytic network,
 which was randomly chosen in this study.
From the results of simulations of a variety of networks,
 we have obtained the steady state in which a cell grows with a constant rate.

We have carried out simulations by varying the values of
 the concentration of the nutrient outside the cell, i.e., $X_0$,
 and the rate constant of the phenomenological growth reaction, i.e., $k_g$.
Results of these simulations show that
 three distinct phases with qualitatively different cellular states exhibiting steady growth exist,
 depending on these two parameters.
When the external nutrient concentration $X_0$ is high,
 phase I appears for a large value of $k_g$;this phase is characterized
 by high cell growth and influx rates.
On the other hand, when the value of $k_g$ is small,
 phase II appears, characterized by a low growth rate and a high influx rate.
At an intermediate value of $k_g$, these two phases coexist.
When $X_0$ is low, the influx rate decreases,
 and this regime where phases I and II coexist narrows and then completely disappears.
After the disappearance of phases I and II,
 phase III comes into existence;this phase is characterized by
 low growth and influx rates, regardless of the values of $k_g$.
Further, this phase is characterized by a high efficiency of conversion of nutrients,
 which enable an increase in the cell volume.
In addition to these states characterized by a positive growth rate,
 we also found phase IV, termed the ``death'' phase; this phase is characterized
 by negative growth rates for low $X_0$.
When $k_g$ is high, it is found that both the death phase and the growth phase I or III coexist.

These transitions between the phases are explained
 using the mean field equations of the model,
 which are obtained by assuming that the path density is sufficiently high.
The mean field equations account well for the transition
 between phases I and II, as well as those between phase I or III and IV
 as saddle-node bifurcations.
The success of the mean field equations that are independent of each network
 also explains why the behaviors observed here
 are independent of the choice of networks.
The analysis of the mean field equations enables us to understand
 the bistability of two phases as
 mutual inhibition between the cell volume growth and the concentration of intracellular chemicals.
Because this mutual inhibition does not depend on the details of the model,
 we expect that this bistability will be a general feature of
 any cells in which the rate of conversion of nutrient for increasing
 cell volume can vary depending on the composition of chemicals in the cell.

In the original simulations of many chemical components,
 we have also examined the statistical distribution of the abundance of chemicals.
To characterize each cellular phase statistically, we adopt the distribution of abundances.
In phase I, the abundance of chemicals differs from species to species, spreading over several orders,
 and this distribution is far from that expected from
 the equilibrium distribution defined by chemical potentials.
As the growth rate increases, the abundance distribution of chemicals
 approaches a certain power law, known as Zipf's law.
On the other hand, the abundance distribution of chemicals in phase II
 is close to the equilibrium distribution.
Hence, the transition from phase II to I is a bifurcation
 from a branch near equilibrium to that far from equilibrium.
It should be noted that the power law with respect to the abundance distribution was also observed
 in the previous model, without consideration of energy conversion\cite{cf2003prl},
 while phases II, III, and IV were first observed in this study.

It is interesting to note that the power law behavior exhibited in the protocell model in this study
 is also commonly observed in the existing cells with regard to gene expression levels\cite{cf2003prl}.
In particular it is confirmed in the log phase in bacteria
where cells grow efficiently as in the phase I in our model.
Note that the power law in abundances in protocell models
is observed for a state in which a cell  grows efficiently keeping its chemical composition.
The distribution of abundances over chemicals
may provide a proper statistical measure for a cellular state.

Because of the simplicity of our model,
 it is not so easy to directly compare our results
 with the behaviors of the existing cells.
However, it is still interesting to note the similarity of the phases observed in this study with
the log, stationary, and dormant phases.
The log phase appears under a nutrient-rich condition,
 where the cell has a high growth rate with a high influx rate.
The stationary phase appears when a nutrient becomes exhausted
 or toxic waste products become excessive, and the cell growth slows.
In the dormant phase, the cell growth rate is suppressed even under a nutrient-rich condition\cite{nqb2004sci}.
These behaviors are reminiscent of phases I, III, and II, respectively.
In this respect, it will be interesting to check whether
 there is hysteresis between the transitions from the log phase to
 the other phases at the single-cell level and to measure the efficiency of cell growth.  Mutual inhibition
between the cell growth and catalyst abundances to support the bistability of the two phases
can be examined directly.
As mentioned above, statistical  distribution of chemical abundances is distinct by each phase,
which may be examined from gene expression analysis.  So far, no reliable microarray data 
are available for stationary and dormant phases, which should be important in the future.

The present model does not assume a specific reaction network
 and intracellular spatial organizations.
Currently, several attempts are being made to construct
 an artificial cell that grows and replicates
\cite{book:rasmussen},
 and the basic structures of the present model
 are often satisfied with such an artificial cell.
Hence, it will be interesting to search for phases by measuring
 the growth rate and the influx rate for cell growth.

In general, there is no established theory for thermodynamics far from equilibrium.
The metabolic efficiency $\eta$ discussed above is also not, in a strict sense,
 the thermodynamic efficiency.
However, by restricting the theory to a steadily growing state,
 we may hope to develop a thermodynamic theory on the basis of which the efficiency
 of cell growth is analyzed\cite{hvw1983pns}.
The present model exhibits three distinct phases with regard to the influx and growth rates,
 with unique metabolic efficiencies.
Such a thermodynamic theory, if established,
 will account for these phases and efficiency of cell growth.

\begin{acknowledgments}
We would like to thank Tetsuya Yomo, Shuji Ishihara, Chikara Furusawa,
 and Stuart Kauffman for illuminating comments and stimulating discussions.
The present work is partially supported by the ERATO "Dynamical Micro-Scale Reaction Environment project"
 by Japan Science and Technology Agency (JST).
We are grateful to Japanese people for generous support.
\end{acknowledgments}

\begin{thebibliography}{10}%
\makeatletter
\providecommand \@ifxundefined [1]{%
 \ifx #1\undefined \expandafter \@firstoftwo
 \else \expandafter \@secondoftwo
\fi
}%
\providecommand \@ifnum [1]{%
 \ifnum #1\expandafter \@firstoftwo
 \else \expandafter \@secondoftwo
\fi
}%
\providecommand \enquote [1]{``#1''}%
\providecommand \bibnamefont  [1]{#1}%
\providecommand \bibfnamefont [1]{#1}%
\providecommand \citenamefont [1]{#1}%
\providecommand\href[0]{\@sanitize\@href}%
\providecommand\@href[1]{\endgroup\@@startlink{#1}\endgroup\@@href}%
\providecommand\@@href[1]{#1\@@endlink}%
\providecommand \@sanitize [0]{\begingroup\catcode`\&12\catcode`\#12\relax}%
\@ifxundefined \pdfoutput {\@firstoftwo}{%
 \@ifnum{\z@=\pdfoutput}{\@firstoftwo}{\@secondoftwo}%
}{%
 \providecommand\@@startlink[1]{\leavevmode}%
 \providecommand\@@endlink[0]{}%
}{%
 \providecommand\@@startlink[1]{%
  \leavevmode
  \pdfstartlink
   attr{/Border[0 0 1 ]/H/I/C[0 1 1]}%
   user{/Subtype/Link/A<</Type/Action/S/URI/URI(#1)>>}%
  \relax
 }%
 \providecommand\@@endlink[0]{\pdfendlink}%
}%
\providecommand \url  [0]{\begingroup\@sanitize \@url }%
\providecommand \@url [1]{\endgroup\@href {#1}{\urlprefix}}%
\providecommand \urlprefix [0]{URL }%
\providecommand \Eprint[0]{\href }%
\@ifxundefined \urlstyle {%
  \providecommand \doi [1]{doi:\discretionary{}{}{}#1}%
}{%
  \providecommand \doi [0]{doi:\discretionary{}{}{}\begingroup
  \urlstyle{rm}\Url }%
}%
\providecommand \doibase [0]{http://dx.doi.org/}%
\providecommand \Doi[1]{\href{\doibase#1}}%
\providecommand \bibAnnote [3]{%
  \BibitemShut{#1}%
  \begin{quotation}\noindent
    \textsc{Key:}\ #2\\\textsc{Annotation:}\ #3%
  \end{quotation}%
}%
\providecommand \bibAnnoteFile [2]{%
  \IfFileExists{#2}{\bibAnnote {#1} {#2} {\input{#2}}}{}%
}%
\providecommand \typeout [0]{\immediate \write \m@ne }%
\providecommand \selectlanguage [0]{\@gobble}%
\providecommand \bibinfo [0]{\@secondoftwo}%
\providecommand \bibfield [0]{\@secondoftwo}%
\providecommand \translation [1]{[#1]}%
\providecommand \BibitemOpen[0]{}%
\providecommand \bibitemStop [0]{}%
\providecommand \bibitemNoStop [0]{.\EOS\space}%
\providecommand \EOS [0]{\spacefactor3000\relax}%
\providecommand \BibitemShut [1]{\csname bibitem#1\endcsname}%
%</preamble>
\bibitem{book:schrodinger}%
  \BibitemOpen
  \bibfield{author}{%
  \bibinfo {author} {\bibfnamefont{E.}~\bibnamefont{Schr\"{o}dinger}},\ }%
  \emph{\bibinfo {title} {What is life?}}\ (\bibinfo {publisher} {Cambridge
  University Press, Cambridge, England},\ \bibinfo {year} {1944})%
  \bibAnnoteFile{NoStop}{book:schrodinger}%
\bibitem{book:prigogine}%
  \BibitemOpen
  \bibfield{author}{%
  \bibinfo {author} {\bibfnamefont{G.}~\bibnamefont{Nicolis}}\ and\ \bibinfo
  {author} {\bibfnamefont{I.}~\bibnamefont{Prigogine}},\ }%
  \emph{\bibinfo {title} {Self-Organization in Nonequilibrium Systems}}\
  (\bibinfo {publisher} {Wiley \& Sons, NewYork},\ \bibinfo {year} {1977})%
  \bibAnnoteFile{NoStop}{book:prigogine}%
\bibitem{jbr1995mr}%
  \BibitemOpen
  \bibfield{author}{%
  \bibinfo {author} {\bibfnamefont{J.~B.}\ \bibnamefont{Russell}}\ and\ \bibinfo
  {author} {\bibfnamefont{G.~M.}\ \bibnamefont{Cook}},\ }%
  \bibfield{journal}{%
  \bibinfo {journal} {Microbiological Revies}\ }%
  \textbf{\bibinfo {volume} {59}},\ \bibinfo {pages} {48} (\bibinfo {year}
  {1995})%
  \bibAnnoteFile{NoStop}{jbr1995mr}%
\bibitem{jm1949arm}%
  \BibitemOpen
  \bibfield{author}{%
  \bibinfo {author} {\bibfnamefont{J.}~\bibnamefont{Monod}},\ }%
  \bibfield{journal}{%
  \bibinfo {journal} {Annu. Rev. Microbiol.}\ }%
  \textbf{\bibinfo {volume} {3}},\ \bibinfo {pages} {371} (\bibinfo {year}
  {1949})%
  \bibAnnoteFile{NoStop}{jm1949arm}%
\bibitem{nqb2004sci}%
  \BibitemOpen
  \bibfield{author}{%
  \bibinfo {author} {\bibfnamefont{N.~Q.}\ \bibnamefont{Balaban}}, \bibinfo
  {author} {\bibfnamefont{J.}~\bibnamefont{Merrin}}, \bibinfo {author}
  {\bibfnamefont{R.}~\bibnamefont{Chait}}, \bibinfo {author}
  {\bibfnamefont{L.}~\bibnamefont{Kowalik}},\ and\ \bibinfo {author}
  {\bibfnamefont{S.}~\bibnamefont{Leibler}},\ }%
  \bibfield{journal}{%
  \bibinfo {journal} {Science}\ }%
  \textbf{\bibinfo {volume} {305}},\ \bibinfo {pages} {1622} (\bibinfo {year}
  {2004})%
  \bibAnnoteFile{NoStop}{nqb2004sci}%
\bibitem{ik2004jb}%
  \BibitemOpen
  \bibfield{author}{%
  \bibinfo {author} {\bibfnamefont{I.}~\bibnamefont{Keren}}, \bibinfo {author}
  {\bibfnamefont{D.}~\bibnamefont{Shah}}, \bibinfo {author}
  {\bibfnamefont{A.}~\bibnamefont{Spoering}}, \bibinfo {author}
  {\bibfnamefont{N.}~\bibnamefont{Kaldalu}},\ and\ \bibinfo {author}
  {\bibfnamefont{K.}~\bibnamefont{Lewis}},\ }%
  \bibfield{journal}{%
  \bibinfo {journal} {Journal of Bacteriology}\ }%
  \textbf{\bibinfo {volume} {186}},\ \bibinfo {pages} {8172} (\bibinfo {year}
  {2004})%
  \bibAnnoteFile{NoStop}{ik2004jb}%
\bibitem{book:hypercycle}%
  \BibitemOpen
  \bibfield{author}{%
  \bibinfo {author} {\bibfnamefont{M.}~\bibnamefont{Eigen}}\ and\ \bibinfo
  {author} {\bibfnamefont{P.}~\bibnamefont{Schuster}},\ }%
  \emph{\bibinfo {title} {The Hypercycle}}\ (\bibinfo {publisher} {Springer},\
  \bibinfo {year} {1979})%
  \bibAnnoteFile{NoStop}{book:hypercycle}%
\bibitem{tg1975bs}%
  \BibitemOpen
  \bibfield{author}{%
  \bibinfo {author} {\bibfnamefont{T.}~\bibnamefont{G\'anti}},\ }%
  \bibfield{journal}{%
  \bibinfo {journal} {BioSystems}\ }%
  \textbf{\bibinfo {volume} {7}},\ \bibinfo {pages} {15} (\bibinfo {year}
  {1975})%
  \bibAnnoteFile{NoStop}{tg1975bs}%
\bibitem{rjb1991al2}%
  \BibitemOpen
  \bibfield{author}{%
  \bibinfo {author} {\bibfnamefont{R.~J.}\ \bibnamefont{Bagley}}\ and\ \bibinfo
  {author} {\bibfnamefont{J.~D.}\ \bibnamefont{Farmer}},\ }%
  \bibfield{journal}{%
  \bibinfo {journal} {in Artificial Life II ed. C.G.Langton et al., Addison Wesley, Redwood City, CA},\ \bibinfo {pages} {93}}%
   (\bibinfo {year} {1991})%
  \bibAnnoteFile{NoStop}{rjb1991al2}%
\bibitem{ds1998oleb}%
  \BibitemOpen
  \bibfield{author}{%
  \bibinfo {author} {\bibfnamefont{D.}~\bibnamefont{Segr\`e}}, \bibinfo
  {author} {\bibfnamefont{D.}~\bibnamefont{Lancet}}, \bibinfo {author}
  {\bibfnamefont{O.}~\bibnamefont{Kedem}},\ and\ \bibinfo {author}
  {\bibfnamefont{Y.}~\bibnamefont{Pilpel}},\ }%
  \bibfield{journal}{%
  \bibinfo {journal} {Origins Life Evol.Biosphere}\ }%
  \textbf{\bibinfo {volume} {28}},\ \bibinfo {pages} {501} (\bibinfo {year}
  {1998})%
  \bibAnnoteFile{NoStop}{ds1998oleb}%
\bibitem{sj2001pns}%
  \BibitemOpen
  \bibfield{author}{%
  \bibinfo {author} {\bibfnamefont{S.}~\bibnamefont{Jain}}\ and\ \bibinfo
  {author} {\bibfnamefont{S.}~\bibnamefont{Krishna}},\ }%
  \bibfield{journal}{%
  \bibinfo {journal} {Proc. Nat. Acad. Sci. USA}\ }%
  \textbf{\bibinfo {volume} {98}},\ \bibinfo {pages} {543} (\bibinfo {year}
  {2001})%
  \bibAnnoteFile{NoStop}{sj2001pns}%
\bibitem{cf2003prl}%
  \BibitemOpen
  \bibfield{author}{%
  \bibinfo {author} {\bibfnamefont{C.}~\bibnamefont{Furusawa}}\ and\ \bibinfo
  {author} {\bibfnamefont{K.}~\bibnamefont{Kaneko}},\ }%
  \bibfield{journal}{%
  \bibinfo {journal} {Phys.Rev.Lett}\ }%
  \textbf{\bibinfo {volume} {90}} (\bibinfo {year} {2003}),\ \bibinfo {note}
  {088102}%
  \bibAnnoteFile{NoStop}{cf2003prl}%
\bibitem{cf2005bp}%
  \BibitemOpen
  \bibfield{author}{%
  \bibinfo {author} {\bibfnamefont{C.}~\bibnamefont{Furusawa}}, \bibinfo
  {author} {\bibfnamefont{T.}~\bibnamefont{Suzuki}}, \bibinfo {author}
  {\bibfnamefont{A.}~\bibnamefont{Kashiwagi}}, \bibinfo {author}
  {\bibfnamefont{T.}~\bibnamefont{Yomo}},\ and\ \bibinfo {author}
  {\bibfnamefont{K.}~\bibnamefont{Kaneko}},\ }%
  \bibfield{journal}{%
  \bibinfo {journal} {BIOPHYSICS}\ }%
  \textbf{\bibinfo {volume} {1}} (\bibinfo {year} {2005}),\ \bibinfo {note}
  {25}%
  \bibAnnoteFile{NoStop}{cf2005bp}%
\bibitem{kk2005acp}%
  \BibitemOpen
  \bibfield{author}{%
  \bibinfo {author} {\bibfnamefont{K.}~\bibnamefont{Kaneko}},\ }%
  \bibfield{journal}{%
  \bibinfo {journal} {Adv.Chem.Phys}\ }%
  \textbf{\bibinfo {volume} {130}},\ \bibinfo {pages} {543} (\bibinfo {year}
  {2005})%
  \bibAnnoteFile{NoStop}{kk2005acp}%
\bibitem{cf2006pre}%
  \BibitemOpen
  \bibfield{author}{%
  \bibinfo {author} {\bibfnamefont{C.}~\bibnamefont{Furusawa}}\ and\ \bibinfo
  {author} {\bibfnamefont{K.}~\bibnamefont{Kaneko}},\ }%
  \bibfield{journal}{%
  \bibinfo {journal} {Phys.Rev E}\ }%
  \textbf{\bibinfo {volume} {73}} (\bibinfo {year} {2006}),\ \bibinfo {note}
  {011912}%
  \bibAnnoteFile{NoStop}{cf2006pre}%
\bibitem{tr2007ptb}%
  \BibitemOpen
  \bibfield{author}{%
  \bibinfo {author} {\bibfnamefont{T.}~\bibnamefont{Rocheleau}}, \bibinfo
  {author} {\bibfnamefont{S.}~\bibnamefont{Rasmussen}}, \bibinfo {author}
  {\bibfnamefont{P.~E.}\ \bibnamefont{Nielsen}}, \bibinfo {author}
  {\bibfnamefont{M.~N.}\ \bibnamefont{Jacobi}},\ and\ \bibinfo {author}
  {\bibfnamefont{H.}~\bibnamefont{Ziock}},\ }%
  \bibfield{journal}{%
  \bibinfo {journal} {Phil. Trans. R. Soc. B}\ }%
  \textbf{\bibinfo {volume} {362}},\ \bibinfo {pages} {1841} (\bibinfo {year}
  {2007})%
  \bibAnnoteFile{NoStop}{tr2007ptb}%
\bibitem{tc2008jtb}%
  \BibitemOpen
  \bibfield{author}{%
  \bibinfo {author} {\bibfnamefont{T.}~\bibnamefont{Carletti}}, \bibinfo
  {author} {\bibfnamefont{R.}~\bibnamefont{Serra}}, \bibinfo {author}
  {\bibfnamefont{I.}~\bibnamefont{Poli}}, \bibinfo {author}
  {\bibfnamefont{M.}~\bibnamefont{Villani}},\ and\ \bibinfo {author}
  {\bibfnamefont{A.}~\bibnamefont{Filisetti}},\ }%
  \bibfield{journal}{%
  \bibinfo {journal} {J. Theor. Biol.}\ }%
  \textbf{\bibinfo {volume} {254}},\ \bibinfo {pages} {741} (\bibinfo {year}
  {2008})%
  \bibAnnoteFile{NoStop}{tc2008jtb}%
\bibitem{me2007bpj}%
  \BibitemOpen
  \bibfield{author}{%
  \bibinfo {author} {\bibfnamefont{M.}~\bibnamefont{Ederer}}\ and\ \bibinfo
  {author} {\bibfnamefont{E.}~\bibnamefont{Gilles}},\ }%
  \bibfield{journal}{%
  \bibinfo {journal} {Biophysical Journal}\ }%
  \textbf{\bibinfo {volume} {92}},\ \bibinfo {pages} {1846} (\bibinfo {year}
  {2007})%
  \bibAnnoteFile{NoStop}{me2007bpj}%
\bibitem{aa2009pre}%
  \BibitemOpen
  \bibfield{author}{%
  \bibinfo {author} {\bibfnamefont{A.}~\bibnamefont{Awazu}}\ and\ \bibinfo
  {author} {\bibfnamefont{K.}~\bibnamefont{Kaneko}},\ }%
  \bibfield{journal}{%
  \bibinfo {journal} {Phys.Rev.E}\ }%
  \textbf{\bibinfo {volume} {80}} (\bibinfo {year} {2009}),\ \bibinfo {note}
  {041931}%
  \bibAnnoteFile{NoStop}{aa2009pre}%
\bibitem{sak1986jtb}%
  \BibitemOpen
  \bibfield{author}{%
  \bibinfo {author} {\bibfnamefont{S.~A.}\ \bibnamefont{Kauffman}},\ }%
  \bibfield{journal}{%
  \bibinfo {journal} {J.Theor.Biol}\ }%
  \textbf{\bibinfo {volume} {119}},\ \bibinfo {pages} {1} (\bibinfo {year}
  {1986})%
  \bibAnnoteFile{NoStop}{sak1986jtb}%
\bibitem{pfs1993phd}%
  \BibitemOpen
  \bibfield{author}{%
  \bibinfo {author} {\bibfnamefont{P.~F.}\ \bibnamefont{Stadler}}, \bibinfo
  {author} {\bibfnamefont{W.}~\bibnamefont{Fontana}},\ and\ \bibinfo {author}
  {\bibfnamefont{J.~H.}\ \bibnamefont{Miller}},\ }%
  \bibfield{journal}{%
  \bibinfo {journal} {Physica D}\ }%
  \textbf{\bibinfo {volume} {63}},\ \bibinfo {pages} {378} (\bibinfo {year}
  {1993})%
  \bibAnnoteFile{NoStop}{pfs1993phd}%
\bibitem{rh2005pre}%
  \BibitemOpen
  \bibfield{author}{%
  \bibinfo {author} {\bibfnamefont{R.}~\bibnamefont{Hanel}}, \bibinfo {author}
  {\bibfnamefont{S.~A.}\ \bibnamefont{Kauffman}},\ and\ \bibinfo {author}
  {\bibfnamefont{S.}~\bibnamefont{Thurner}},\ }%
  \bibfield{journal}{%
  \bibinfo {journal} {Phys. Rev. E}\ }%
  \textbf{\bibinfo {volume} {72}} (\bibinfo {year} {2005}),\ \bibinfo {note}
  {036117}%
  \bibAnnoteFile{NoStop}{rh2005pre}%
\bibitem{es2008jtb}%
  \BibitemOpen
  \bibfield{author}{%
  \bibinfo {author} {\bibfnamefont{E.}~\bibnamefont{Smith}},\ }%
  \bibfield{journal}{%
  \bibinfo {journal} {J.Theor.Biol}\ }%
  \textbf{\bibinfo {volume} {252}},\ \bibinfo {pages} {198} (\bibinfo {year}
  {2008})%
  \bibAnnoteFile{NoStop}{es2008jtb}%
\bibitem{book:rasmussen}%
  \BibitemOpen
  \emph{\bibinfo {title} {Protocells:Bridging Nonliving and Living Matter}},\
  edited by\ \bibinfo {editor} {\bibfnamefont{S.}~\bibnamefont{Rasmussen}},
  \bibinfo {editor} {\bibfnamefont{M.~A.}\ \bibnamefont{Bedau}}, \bibinfo
  {editor} {\bibfnamefont{L.}~\bibnamefont{Chen}}, \bibinfo {editor}
  {\bibfnamefont{D.}~\bibnamefont{Deamer}}, \bibinfo {editor}
  {\bibfnamefont{D.~C.}\ \bibnamefont{Krakauer}}, \bibinfo {editor}
  {\bibfnamefont{N.~H.}\ \bibnamefont{Packard}},\ and\ \bibinfo {editor}
  {\bibfnamefont{P.~F.}\ \bibnamefont{Stadler}}\ (\bibinfo {publisher} {MIT
  Press, Cambridge, MA},\ \bibinfo {year} {2009})%
  \bibAnnoteFile{NoStop}{book:rasmussen}%
\bibitem{hvw1983pns}%
  \BibitemOpen
  \bibfield{author}{%
  \bibinfo {author} {\bibfnamefont{H.~V.}\ \bibnamefont{Westerhoff}}, \bibinfo
  {author} {\bibfnamefont{K.~J.}\ \bibnamefont{Hellingwerf}},\ and\ \bibinfo
  {author} {\bibfnamefont{K.~V.}\ \bibnamefont{Dam}},\ }%
  \bibfield{journal}{%
  \bibinfo {journal} {Proc. Natl. Acad. Sci. USA}\ }%
  \textbf{\bibinfo {volume} {80}},\ \bibinfo {pages} {305} (\bibinfo {year}
  {1983})%
  \bibAnnoteFile{NoStop}{hvw1983pns}%
\end{thebibliography}
\end{document}